\documentclass[aps,amssymb,showpacs]{revtex4}
\usepackage{amssymb}
\usepackage{graphicx}
\usepackage{color}
\usepackage{textcomp}
\usepackage{ulem}
\begin{document}

\title{ Pions in magnetic field at finite temperature }
\author{Shijun Mao}
\affiliation{School of Science, Xi'an Jiaotong University, Xi'an, Shaanxi 710049, China}

\begin{abstract}
Pions in external magnetic field are investigated in the frame of a Pauli-Villars regularized Nambu--Jona-Lasinio model. The meson propagators in terms of quark bubbles in Ritus and Schwinger schemes are analytically derived, and pion masses are numerically calculated in the Ritus scheme. For neutral and charged pions at finite temperature, there exist respectively one and three mass jumps at the corresponding Mott transition points, due to the discrete energy levels of the two constituent quarks in magnetic field.
\end{abstract}

\date{\today}
\maketitle
\section{Introduction}
The electromagnetic field effect on Quantum Chromodynamics (QCD) matter draws much attention in recent years, due to its close relation to high energy nuclear collisions~\cite{hi0,hi1,hi2,hi3,hi31,hi32,hi4,hi5,hi6}, compact stars~\cite{ns0,ns1,ns2,ns21,ns3,ns4,ns41,ns42,ns43,ns5,ns6} and cosmological phase transitions~\cite{cpt1,cpt2,cpt3,cpt4}. There appear many interesting topics, such as inverse magnetic catalysis in chiral and deconfinement phase transitions~\cite{lattice1,lattice2,lattice3,lattice4,rev1,rev2,rev3,rev4}, charged vector meson condensation in vacuum~\cite{rho1,l1}, and chiral magnetic effect induced by nontrivial topological structure of QCD~\cite{hi0,hi1,hi2,hi3}.

An important question in the study of electromagnetic fields coupled to strong interaction is the change in hadron properties. The hadron properties in external electromagnetic fields depend on whether we consider the hadron inner structure. For instance, for neutral hadrons, they do not have any interaction with electromagnetic fields, if we neglect the charged constituent quarks. Meson properties in electromagnetic fields are widely studied in effective models at hadron level, such as in chiral perturbation theory~\cite{c1,c3,hadron1} and linear sigma model~\cite{hadron2}, and at quark level, such as in Nambu--Jona-Lasinio (NJL) model~\cite{njl2,meson,mfir,ritus5,ritus6,phi,maopion,wang,liuhao1,liuhao2,liuhao3,coppola} and lattice QCD simulations~\cite{l1,l2,l3,l4}.

In the NJL model at quark level, mesons are treated as quantum fluctuations above the mean field for quarks. They are usually introduced through random phase approximation (RPA) or bosonization~\cite{njl1,njl2,njl3,njl4,njl5,zhuang}. Under bosonization, the quark system in the original NJL model is changed into a meson system, which works well in the low temperature region. The advantage of the RPA method is its application at both low and high temperatures~\cite{zhuang}, since the RPA method leads to a quark-meson system, where quarks dominate the thermodynamics at high temperature and mesons control the system at low temperature. The neutral mesons ($\pi_0, \sigma$) in magnetized NJL model are investigated by taking different methods~\cite{meson,mfir,ritus5,ritus6,phi,maopion,wang,coppola}, all giving consistent results with the lattice QCD in low temperature limit~\cite{l1,l2,l3,l4}. However, the charged mesons, like $\pi_\pm$ and $\rho_\pm$, have not yet been satisfactorily investigated. The difficulty lies in the lack of translational invariance, when external electromagnetic fields are turned on. As a result, the Fourier transformation between coordinate and momentum spaces is not as simple as for neutral mesons. In some of the calculations, charged mesons are studied by simply assuming translational invariance~\cite{phi,liuhao1,liuhao2,liuhao3}. Recently, $\pi_{\pm}$ mesons are constructed in a magnetized NJL model through bosonization~\cite{wang}, obtaining reasonable pion masses at low temperature but failing at high temperature. Charged pions in vacuum are also studied~\cite{coppola} by employing the Ritus eigenfunction method. In this paper, by seriously taking into account the breaking of translational invariance, we investigate at finite temperature $\pi_\pm$ mesons under an external magnetic field through the RPA method in the NJL model.

\section{Formalism}
The SU(2) NJL model is defined through the Lagrangian density in terms of quark fields $\psi$~\cite{njl1,njl2,njl3,njl4,njl5}
\begin{equation}
\label{njl}
{\cal L} = \bar{\psi}\left(i\gamma_\mu D^\mu-m_0\right)\psi+G \left[\left(\bar\psi\psi\right)^2+\left(\bar\psi i\gamma_5{\vec \tau}\psi\right)^2\right],
\end{equation}
where the covariant derivative $D_\mu=\partial_\mu+iQ A_\mu$ couples quarks with electric charge $Q=diag (Q_u,Q_d)=diag (2e/3,-e/3)$ to the external magnetic field ${\bf B}=(0, 0, B)$ in $z$-direction through the potential $A_\mu=(0,0,Bx_1,0)$, $G$ is the coupling constant in scalar and pseudo-scalar channels, and $m_0$ is the current quark mass characterizing the explicit chiral symmetry breaking. We use throughout the paper the definition $x^\mu=(x_0,x_1,x_2,x_3)$ and $p^\mu=(p_0,p_1,p_2,p_3)$.

In mean field approximation, the thermodynamic potential of the system at finite temperature $T$ and external magnetic field $B$ contains the mean field part and the quark part,
\begin{eqnarray}
\Omega_{mf} &=& \frac{(m_q-m_0)^2}{4 G}+\Omega_q,\nonumber\\
\Omega_q &=& -N_c \sum_{f,n}\alpha_n  \frac{|Q_f B|}{2\pi} \int \frac{d p_3}{2\pi}\left[E_f+ 2T \ln \left(1+e^{-E_f/T}\right)\right]
\end{eqnarray}
with the summation over all flavors and Landau energy levels, spin factor $\alpha_n=2-\delta_{n0}$, quark energy $E_f=\sqrt{p^2_3+2 n |Q_f B|+m_q^2}$, and the number of colors $N_c=3$ which is trivial in the NJL model. The chiral condensate $\langle\bar\psi\psi\rangle$ or the dynamical quark mass $m_q=m_0-2G\langle\bar\psi\psi\rangle$ is controlled by the minimum of the thermodynamic potential~\cite{rev1,rev2,rev3,rev4},
\begin{equation}
\label{gap}
{\partial\Omega_{mf}\over \partial m_q}=m_q(1-2GJ_1)-m_0=0
\end{equation}
with
\begin{equation}
J_1 = N_c\sum_{f,n}\alpha_n \frac{|Q_f B|}{2\pi} \int \frac{d p_3}{2\pi} \frac{\tanh\left(\frac{E_f}{2T}\right)}{ E_f}.
\end{equation}

As quantum fluctuations above the mean field, mesons are constructed through quark bubble summation in the frame of RPA~\cite{njl2,njl3,njl4,njl5,zhuang}. Namely, the quark interaction via a meson exchange is effectively described by using the Dyson-Schwinger equation,
\begin{equation}
{\cal D}_M(x,z)  = 2G \delta(x-z)+\int d^4y\ 2G \Pi_M(x,y) {\cal D}_M(y,z),
\label{dsequ}
\end{equation}
where ${\cal D}_M(x,y)$ represents the meson propagator from $x$ to $y$, and the corresponding meson polarization function is the quark bubble,
\begin{equation}
\label{bubble}
\Pi_M(x,y) = i{\text {Tr}}\left[\Gamma_M^{\ast} S(x,y) \Gamma_M  S(y,x)\right]
\end{equation}
with the meson vertex
\begin{equation}
\label{vertex} \Gamma_M = \left\{\begin{array}{ll}
1 & M=\sigma\\
i\tau_+\gamma_5 & M=\pi_+ \\
i\tau_-\gamma_5 & M=\pi_- \\
i\tau_3\gamma_5 & M=\pi_0\ ,
\end{array}\right.
\Gamma_M^* = \left\{\begin{array}{ll}
1 & M=\sigma\\
i\tau_-\gamma_5 & M=\pi_+ \\
i\tau_+\gamma_5 & M=\pi_- \\
i\tau_3\gamma_5 & M=\pi_0\ .
\end{array}\right. \nonumber
\end{equation}
The quark propagator matrix in flavor space $S=diag(S_u,\ S_d)$ is at mean field level, and the trace is taken in spin, color and flavor spaces.

There are two equivalent ways to treat the quark and meson propagators in magnetic field, the Schwinger scheme~\cite{rev3,rev4} and the Ritus scheme~\cite{ritus1,ritus2,ritus21,ritus3,ritus4,ritus7}. In the following, we derive the meson pole equations with the two methods and make comparison between them.

\subsection{Ritus scheme}
In Ritus scheme, one can well-define the Fourier-like transformation for the particle propagator from the conserved Ritus momentum space to coordinate space~\cite{ritus1,ritus2,ritus21,ritus3,ritus4,ritus7}. The quark propagator with flavor $f$ in coordinate space can be written as
\begin{eqnarray}
\label{quark}
S_f(x,y) &=& \sum_n \int {d^3\tilde p\over (2\pi)^3} e^{-i \tilde p\cdot (x-y)}P_n(x_1,p_2)D_f(\bar p) P_n(y_1,p_2),\nonumber\\
P_n({x_1},p_2) &=& {1\over 2}\left[g_n^{s_f}({x_1},p_2)+I_n g_{n-1}^{s_f}({x_1},p_2)\right]+{is_f\over 2}\left[g_n^{s_f}({x_1},p_2)- I_n g_{n-1}^{s_f}({x_1},p_2)\right]\gamma_1 \gamma_2,\nonumber\\
D_f^{-1}(\bar p) &=& \gamma \cdot \bar p-m_q,
\label{ritusp}
\end{eqnarray}
where $\tilde p=(p_0,0,p_2,p_3)$ is the Fourier transformed momentum, $\bar p=(p_0,0,-s_f \sqrt{2n|Q_f B|},p_3)$ is the conserved Ritus momentum with $n$ describing the quark Landau level in magnetic field, $s_f=\text{sgn}(Q_f B)$ is the quark sign factor, the magnetic field dependent function $g_n^{s_f}({x_1},p_2) = \phi_n(x_1-s_f p_2 /|Q_f B|)$ is controlled by the Hermite polynomial $H_n(\zeta)$ via $\phi_n(\zeta) = \left(2^n n! \sqrt{\pi} |Q_f B|^{-1/2}\right)^{-1/2} e^{-\zeta^2|Q_f B|/2} H_n\left(\zeta/|Q_f B|^{-1/2}\right)$, and $I_n=1-\delta_{n0}$ is governed by the Landau energy level.

In order to illustrate the Ritus scheme clearly, let's first use neutral mesons as a simple example. In this case, the meson momentum $k=(\omega, k_1,k_2,k_3)$ itself is conserved, and the meson propagator and the corresponding meson polarization function in momentum space are just the normal Fourier transformation of their expressions in coordinate space,
\begin{eqnarray}
\label{fourier1}
{\cal D}_{M}(k) &=&  \int d^4 (x-y) e^{i k \cdot (x-y)} {\cal D}_{M}(x,y),\nonumber\\
\Pi_{M}(k) &=&  \int d^4 (x-y) e^{i k \cdot (x-y)} \Pi_{M}(x,y)
\end{eqnarray}
for the pseudo-Goldstone mode $M=\pi_0$ and Higgs mode $M=\sigma$. By taking the quark bubble summation in RPA and considering the complete and orthogonal conditions of the plane wave $e^{-i k \cdot x}$, the meson propagator in momentum space can be simplified as
\begin{equation}
\label{npole}
{\cal D}_M(k)=\frac{2G}{1-2G\Pi_M(k)}.
\end{equation}

The meson pole mass $m_M$ is defined as the pole of the propagator at zero momentum ${\bf k}={\bf 0}$,
\begin{equation}
\label{mmass}
1-2G\Pi_M(\omega=m_M, {\bf 0})=0,
\end{equation}
where the polarization function is greatly simplified at ${\bf k}={\bf 0}$,
\begin{eqnarray}
\label{pi}
\Pi_M(\omega,{\bf 0}) &=& J_1-(\omega^2-\epsilon_M^2) J_2(\omega^2),\\
J_2(\omega^2) &=& -N_c\sum_{f,n}\alpha_n \frac{|Q_f B|}{2\pi} \int \frac{d p_3}{2\pi}{{\tanh\left(\frac{E_f}{2T}\right)}\over  E_f (4 E_f^2-w^2)}\nonumber
\end{eqnarray}
with $\epsilon_{\pi_0}=0$ and $\epsilon_\sigma=2m_q$.

In chiral limit with zero current quark mass $m_0=0$, by comparing the gap equation (\ref{gap}) for mean field quark mass $m_q$ with the pole equation (\ref{mmass}) for neutral meson mass $m_M$, we have the simple relation in the chiral symmetry breaking phase,
\begin{equation}
m_{\pi_0} = 0,\ \ \ \ \ m_\sigma = 2m_q.
\end{equation}
This confirms that $\pi_0$ is the Goldstone mode corresponding to the spontaneous chiral symmetry breaking in external magnetic field.

The above discussed chiral condensate $\langle\bar\psi\psi\rangle$ and the pseudo-Goldstone mode $\pi_0$ and Higgs mode $\sigma$ are charge neutral, and they are affected by the external magnetic field only through the pair of charged quarks. As a consequence, the transformation from coordinate space to momentum space is a conventional Fourier transformation, characterized by the plane wave $e^{-ik x}$. In this case, the equations determining the condensate and meson masses are formally the same as that at ${\bf B}=0$~\cite{zhuang}, except for a replacement of the quark momentum integration $2N_f\int d^3{\bf p}/(2\pi)^3$ by $\sum_{f,n}\alpha_n |Q_f B|/(2\pi)\int dp_3/(2\pi)$.

We now turn to charged mesons. In Ritus scheme~\cite{ritus1,ritus2,ritus21,ritus3,ritus4,ritus7}, the Fourier transformation for neutral mesons (\ref{fourier1}) is extended to
\begin{eqnarray}
\label{fourier2}
{\cal D}_M(\bar k) &=& \int d^4x d^4y F^\ast_k(x){\cal D}_M(x,y) F_k(y),\nonumber\\
\Pi_M(\bar k) &=& \int d^4x d^4y F^\ast_k(x)\Pi_M(x,y) F_k(y)
\label{ritubp}
\end{eqnarray}
for charged mesons $M=\pi_\pm$ with spin zero, where $F_k(x)=e^{-i \tilde k\cdot x} g^{s_M}_l(x_1,k_2)$ is the solution of the Klein-Gordon equation with the index $l$ describing meson Landau level in magnetic field, the Fourier transformed momentum $\tilde k=(k_0,0,k_2,k_3)$ and the meson sign factor $s_M=\text{sgn}(Q_M B)$, and ${\bar k}=(k_0,0,-s_M \sqrt{(2l+1)|Q_M B|},k_3)$ is the conserved four-dimensional Ritus momentum.

Taking into account the complete and orthogonal conditions
\begin{eqnarray}
&&\sum_l \int {d^3\tilde k\over (2\pi)^3}F_k(x)F^\ast_k(y)=\delta^{(4)}(x-y),\\
&& \int d^4x F_k(x) F^\ast_{k'}(x)=(2\pi)^4 \delta^{(3)}({\tilde k}-{\tilde k'}) \delta_{ll'},
\end{eqnarray}
and the Dyson-Schwinger equation (\ref{dsequ}), the meson propagator in momentum space can be simplified as
\begin{equation}
{\cal D}_M(\bar k)=\frac{2G}{1-2G \Pi_M(\bar k)}.
\label{mb}
\end{equation}

Obviously, the propagator for charged mesons has the same structure as for neutral mesons, the only difference lies in the momentum $k$ for $\pi_0$ and $\sigma$ and ${\bar k}$ for $\pi_\pm$. For neutral mesons which do not interact with the magnetic field as a whole object, the momentum $k=(\omega, {\bf k})$ itself is conserved, and the solution of the Klein-Gordon equation is just the plane wave. For charged mesons, however, the solution of the magnetized Klein-Gordon equation is $F_k(x)$ and $\bar k$ is conserved.

The pole mass $m_M$ for charged mesons is defined through the singularity of the meson propagator ${\cal D}^{-1}_M(\bar k)=0$ at $k_0=m_M, l=0$ and $k_3=0$,
\begin{equation}
\label{mpmass}
1-2G\Pi_M(m_M,0,-s_M\sqrt{|Q_M B|},0)=0.
\end{equation}

We now calculate the charged meson polarization function in Ritus-momentum space. Taking the mean field quark propagator (\ref{quark}) and the definition (\ref{bubble}) for the quark bubble, we have the $\pi_+$ polarization function
\begin{eqnarray}
\Pi_{\pi_+}(x,y) &=& 2 i^3 N_c {\text {Tr}}_D \left[\gamma_5 S_u(x,y) \gamma_5 S_d(y,x) \right]\nonumber\\
           &=& 2 i^3 N_c \sum_{n,n'}\int {d^3\tilde p d^3 \tilde q\over (2\pi)^6} e^{-i(\widetilde {p}-\widetilde {q})(x-y)}\nonumber\\
           &\times& {\text {Tr}}_D \left[P_n(x_1,q_2) P_{n'}(x_1,p_2)\frac{-\gamma \cdot {\bar p}+m_q}{{\bar p}^2-m_q^2}+P_n(y_1,q_2)P_{n'}(y_1,p_2)\frac{\gamma \cdot {\bar q}+m_q}{{\bar q}^2-m_q^2} \right]
\end{eqnarray}
in coordinate space and
\begin{eqnarray}
\Pi_{\pi_+}({\bar k}) &=& 8i^3 N_c\int d^4x d^4y\sum_{n,n'}\int {d^3\tilde p d^3 \tilde q\over (2\pi)^6} e^{-i(\widetilde {p}-\widetilde {q} -\widetilde {k})(x-y)}\nonumber\\
                &\times& g_l^+(x_1,k_2) g_l^+(y_1,k_2)\left[A_+(x_1,y_1,k_2) \frac{m_q^2-{\bar p}_0 {\bar q_0}+{\bar p}_3 {\bar q_3}}{({\bar p}^2-m_q^2)({\bar q}^2-m_q^2)}
                         +A_-(x_1,y_1,k_2) \frac{{\bar p}_2 {\bar q_2}}{({\bar p}^2-m_q^2)({\bar q}^2-m_q^2)} \right]
\end{eqnarray}
in Ritus-momentum space with
\begin{eqnarray}
A_\pm(x_1,y_1,k_2)&=& \alpha_+(x_1,k_2)\alpha_+(y_1,k_2)\pm\alpha_-(x_1,k_2)\alpha_-(y_1,k_2), \nonumber\\
\alpha_\pm(z,k_2)&=&\frac{1}{2} \left[I_n g_{n-1}^{s_d}(z,q_2) g_{n'}^{s_u}(z,p_2)\pm u\leftrightarrow d \right].
\end{eqnarray}

Doing the integrations over $x_0,x_2,x_3$ and $y_0,y_2,y_3$, we obtain the momentum conservation relation $\tilde p=\tilde q+\tilde k$. At finite temperature, the integration over the particle energy is replaced by the Fermion/Boson Matsubara frequency summation in the imaginary time formalism of finite temperature field theory. After a straightforward derivation, we have the $\pi_+$ polarization function at the pole,
\begin{eqnarray}
\label{pipm}
\Pi_{\pi^+}(k_0,0,-\sqrt{|eB|},0) &=& J_1+J_3(k^2_0),\nonumber\\
J_3(k^2_0) &=& \sum_{n,n'} \int \frac{d q_3}{2\pi}\frac{j_{n,n'}(k_0^2)}{4E_n E_{n'}}\left[\frac{f(-E_{n'})- f(E_n)}{k_0+E_{n'}+E_n}+\frac{f(E_{n'})- f(-E_n)}{k_0-E_{n'}-E_n}\right],\nonumber\\
j_{n,n'}(k_0^2) &=& \left({k^2_0/2}-n'|Q_u B|-n|Q_d B|\right)j^+_{n,n'} -2 \sqrt{n'|Q_u B|n|Q_d B|}\ j^-_{n,n'},\nonumber\\
j_{n,n'}^\pm &=& \int dx_1dy_1 {d q_2}/{2\pi} A_\pm(x_1,y_1,k_2),
\end{eqnarray}
with the quark energies $E_{n'}=\sqrt{p^2_3+2 n' |Q_u B|+m_q^2}$ and $E_n=\sqrt{q^2_3+2 n |Q_d B|+m_q^2}$ and Fermi-Dirac distribution function $f(E)=1/(e^{E/T}+1)$.  It is easy to prove that the integration over $x_1, y_1$ and $q_2$ is independent of $k_2$. Since the spins of $u$ and ${\bar d}$ quarks at the lowest-Landau-level are always aligned parallel to the external magnetic field, and $\pi_+$ meson has spin zero, the lowest-Landau-level terms with $n=n'=0$ do not contribute to the polarization function, described by $j^\pm_{0,0}=0$.

Considering that the charged pions $\pi_\pm$ with zero spin are Hermite conjugation to each other, they have the same mass even under external magnetic field. In chiral limit, by comparing the quark gap equation (\ref{gap}) with the pole equation (\ref{mpmass}) for $\pi_+$, it is clear that $k_0=m_{\pi_+}=0$ is not the $\pi_+$ pole mass. Under external magnetic field, the $SU(2) \otimes SU(2)$ chiral symmetry is explicitly reduced to $U(1) \otimes U(1)$, charged pions $\pi_{\pm}$ are no longer the Goldstone modes of spontaneous chiral symmetry breaking.

\subsection{Schwinger scheme}
In Schwinger scheme, the particle propagator in magnetic field is composed of two parts, one is related to the Schwinger phase which breaks the translational invariance, and the other is a Fourier transformation of the translational invariant propagator. For instance, the quark propagator with flavor $f$ can be written as~\cite{rev3,rev4} 
\begin{equation}
S_f(x,y) = e^{i \Phi_f(x_{\perp},y_{\perp})} \int {d^4 p\over (2\pi)^4} e^{-ip(x-y)} \widetilde{S}_f(p_{\perp},p_{\parallel})
\label{schwp}
\end{equation}
with the parallel and perpendicular components of the four-dimensional coordinate and momentum, $x_{{\perp}}=\left(x_1,x_2 \right)$, $p_{{\perp}}=\left(p_1,p_2 \right)$ and $p_{\parallel}=\left(p_0,p_3 \right)$. The quark Schwinger phase $\Phi_f$ is defined as
\begin{equation}
\Phi_f(x_{\perp},y_{\perp})=s_f \frac{(x_1+y_1)(x_2-y_2)}{2l_f^2}
\end{equation}
with $l_f=|Q_f B|^{-1/2}$. The translational invariant propagator $\widetilde{S}_f(p_{\perp},p_{\parallel})$ can be decomposed into contributions from different Landau levels,
\begin{equation}
\widetilde{S}_f(p_{\perp},p_{\parallel})=e^{-p^2_{\perp} l_f^2} \sum_n\frac{(-1)^n D_n(p_{\perp},p_{\parallel})}{p^2_{\parallel}-m_q^2-2n|Q_f B|}
\end{equation}
with
\begin{equation}
D_n(p_{\perp},p_{\parallel})= 4\left(\gamma_\perp\cdot p_\perp\right) L^1_{n-1}(2p^2_{\perp} l_f^2)+\left(\gamma^0 p_0 -\gamma^3 p_3+m_q\right)\left(s_- L_n(2p^2_{\perp} l_f^2)-s_+ L_{n-1}(2p^2_{\perp} l_f^2)\right)
\end{equation}
and $s_{\pm}=(1\pm i\gamma_1 \gamma_2 s_f)$, where $L_n(z)$ and $L^a_n(z)$ are Laguerre and generalized Laguerre polynomials.

Similarly, the meson propagator with spin zero can be generally written as~\cite{rev3,rev4},
\begin{equation}
{\cal{D}}_M(x,y)=e^{i \Phi_M(x_{\perp},y_{\perp})} \int {d^4 k\over (2\pi)^4} e^{-ik(x-y)} \widetilde{{\cal{D}}}_M(k_{\perp},k_{\parallel})
\label{schwpb}
\end{equation}
with the meson Schwinger phase $\Phi_M(x_{\perp},y_{\perp})=s_{M} (x_1+y_1)(x_2-y_2)/(2l_M^2)$, $l_M=|Q_M B|^{-1/2}$ and the translational invariant meson propagator $\widetilde{{\cal{D}}}_M(k_{\perp},k_{\parallel})$.

Substituting the quark propagator (\ref{schwp}) and meson propagator (\ref{schwpb}) into the Dyson-Schwinger equation (\ref{dsequ}), and doing the integration over $y$, the $\pi_+$ propagator satisfies the relation,
\begin{eqnarray}
&&\int {d^4 k\over (2\pi)^4} e^{-ik(x-z)} \widetilde{{\cal D}}_{\pi_+}(k_{\perp},k_{\parallel})\nonumber\\
= && 2G\delta(x-z)+4i^3N_cG \int {d^4pd^4q\over (2\pi)^8} e^{-i(q-p)(x-z)}  \widetilde{\cal {D}}_{\pi_+}(k'_{\perp},k'_{\parallel}) {\text {Tr}}_D \left[ \gamma^5 \widetilde{S}_u(q_{\perp},q_{\parallel}) \gamma^5 \widetilde{S}_d(p_{\perp},p_{\parallel}) \right]
\label{mesond}
\end{eqnarray}
with the new meson momentum $k'=(k'_0,k'_1,k'_2,k'_3), k'_0=q_0-p_0, k'_1 =-|eB|(x_2-z_2)/2+(q_1-p_1), k'_2 = -|eB|(z_1-x_1)/2+(q_2-p_2)$ and $k'_3 = q_3-p_3$. This is an integral equation for charged meson propagator. The dependence of the new momentum $k'$ on the coordinate $x-z$ is a direct result of the nontrivial Schwinger phase for charged mesons.

For neutral mesons $\pi_0$ and $\sigma$, we simply have $k'=k=q-p$ and the meson Schwinger phase is automatically cancelled. Therefore, the propagator becomes
\begin{eqnarray}
\widetilde{{\cal D}}_M(k) &=& \frac{2G}{1-2G \widetilde{\Pi}_M(k)},\nonumber\\
\widetilde{\Pi}_M({k}) &=& i^3N_c \sum_f\int {d^4p\over (2\pi)^4} {\text {Tr}}_D \left[ \gamma^5 \widetilde{S}_f(p+k) \gamma^5 \widetilde{S}_f(p) \right].
\end{eqnarray}
It is easy to prove that the polarization function at the pole is exactly the same as (\ref{pi}) in Ritus scheme.

In the end of this section, let's briefly compare the Ritus and Schwinger schemes in constructing mesons under external magnetic field. The difference is the treatment of the breaking of translational symmetry by the magnetic field. A conserved momentum, called Ritus momentum, is introduced in the Ritus scheme and a Schwinger phase is embedded in the particle propagator in the Schwinger scheme. For neutral mesons, the Ritus momentum is reduced to the normal momentum and the Schwinger phase disappears automatically, the two schemes are both convenient and give the same analytic formula. For charged mesons, we still obtain an algebraic equation for the meson propagator in terms of the conserved Ritus momentum, while the nontrivial Schwinger phase leads to an integral equation for the meson propagator which is more difficult to extract the meson properties. In the following, we make numerical calculations in Ritus scheme.

\section{numerical results}
Because of the four-fermion interaction, the NJL model is not a renormalizable theory and needs regularization. The magnetic field does not cause extra ultraviolet divergence but introduces discrete Landau levels and anisotropy in momentum space. To guarantee the law of causality in anisotropic systems, we take into account the Pauli-Villars regularization scheme~\cite{mao}. The three parameters in the NJL model, namely the current quark mass $m_0=5$ MeV, the coupling constant $G=3.44$ GeV$^{-2}$ and the momentum cutoff parameter $\Lambda=1127$ MeV in the Pauli-Villars regularization are fixed by fitting the chiral condensate $\langle\bar\psi\psi\rangle=-(250\ \text{MeV})^3$, pion mass $m_\pi=134$ MeV and pion decay constant $f_\pi=93$ MeV in vacuum at $T=B=0$.

\begin{figure}[hbt]
\centering
\includegraphics[width=7cm]{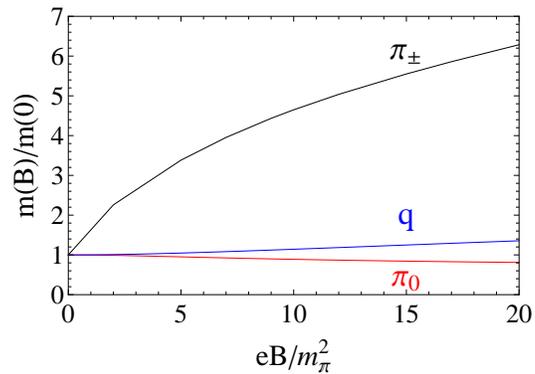}
\caption{The scaled pion and quark masses as functions of magnetic field at $T=0$.} \label{fig1}
\end{figure}

The magnetic field dependence of the pion masses at zero temperature is shown in Fig.\ref{fig1}. As the pseudo-Goldstone mode, the $\pi_0$ mass is a monotonically decreasing function of the magnetic field, $m^2_{\pi_0}(B)\simeq m^2_{\pi}-0.02|eB|$. Due to the isospin symmetry breaking by the external magnetic field, charged pions $\pi_\pm$ are no longer the pseudo-Goldstone modes and become heavier than the $\pi_0$ meson. The $\pi_\pm$ mass increases fast with the magnetic field, $m^2_{\pi_\pm}(B)\simeq m^2_{\pi}+2|eB|$. Our calculations here are qualitatively consistent with the results of other NJL models~\cite{njl2,meson,mfir,ritus5,ritus6,maopion,wang,coppola}, hadron models~\cite{hadron1,c1,c3,hadron2} and lattice QCD simulations~\cite{l1,l2,l3,l4}.

\begin{figure}[hbt]
\centering
\includegraphics[width=11cm]{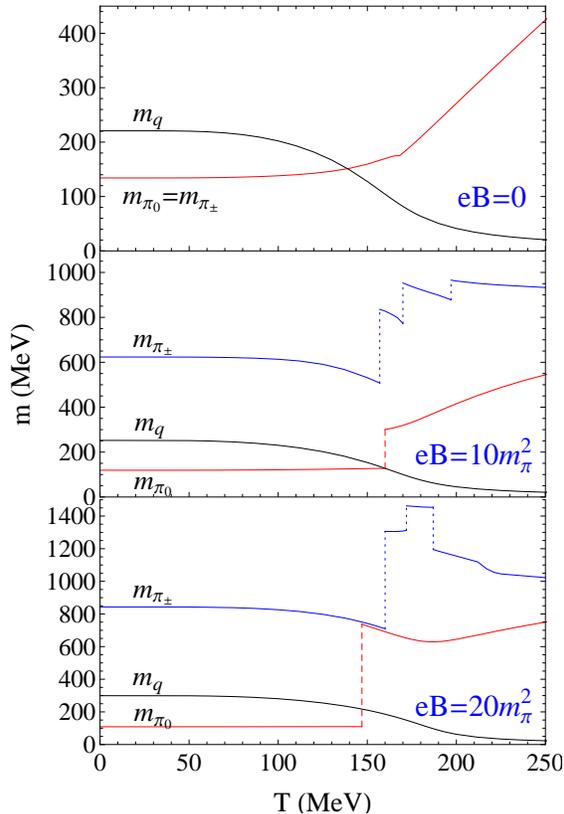}
\caption{The quark and pion masses as functions of temperature at different magnetic field.} \label{fig2}
\end{figure}

The temperature dependence of the quark and pion masses is plotted in Fig.\ref{fig2} at different magnetic field $eB/m^2_\pi=0,10,20$. The dynamical quark mass is gradually melted with increasing temperature, indicating the chiral symmetry restoration. As we discussed in Ref.\cite{maopion}, the pseudo-Goldstone mode $\pi_0$ is controlled by the chiral symmetry, and the mass increases in both symmetry breaking phase at low temperature and symmetry restoration phase at high temperature. When the magnetic field is turned on, there appears a mass jump at the Mott transition temperature $T_m$ where $m_{\pi_0}$ suddenly jumps up from $m_{\pi_0}<2m_q$ to $m_{\pi_0}>2m_q$. Looking back to the pole equation (\ref{mmass}), the jump arises from the infrared singularity of the polarization function $\Pi_{\pi_0}(2m_q,{\bf 0})$ at the lowest Landau level ($n=0$). From our numerical calculation, we have the pseudo critical temperature of chiral symmetry restoration, which is defined by the maximum change of quark mass, $T_c=156, 164, 180$ MeV and the Mott transition temperature $T_m=167, 160, 147$ MeV at magnetic field $eB/m_\pi^2=0,10,20$.

At finite magnetic field and finite temperature, the charged pions $\pi_\pm$ behave totally different from the neutral pion $\pi_0$. The mass $m_{\pi_\pm}$ and $m_{\pi_0}$ approach to each other only in high temperature limit where the finite magnetic field effect is strongly suppressed by the thermal motion, and the translational invariance and isospin symmetry are gradually recovered. Unlike the one mass jump at the Mott transition temperature $T_m$ for the pseudo Goldstone mode, there are three jumps for the charged meson mass $m_{\pi_\pm}$. The difference comes from the structure of the meson polarization function $\Pi_M$ at the pole. For neutral mesons, the constituent quark and antiquark carry the same amount of charge and have the same energy $E_f=\sqrt{p_3^2+2n|Q_f B|+m_q^2}$ and the same mass $E_f(p_3=n=0)=m_q$. Therefore, the Mott transition point is located at $m_{\pi_0}=2m_q$. For charged mesons, however, the constituent quark and antiquark carry different amount of charge and have different energies $E_u=\sqrt{p_3^2+2n'|Q_uB|+m_q^2}$ and $E_d=\sqrt{p_3^2+2n|Q_dB|+m_q^2}$. As we analyzed in the last section, the two constituents at the lowest Landau level ($n=n'=0$) do not satisfy the pole equation (\ref{mpmass}), namely they can not form a charged meson. In this case, the charged meson decay $\pi_+\to u+\bar d$ happens at three threshold energies, $m_{\pi_+}=m_q+\sqrt{2{|Q_uB|}+m_q^2}, m_q+\sqrt{2{|Q_dB|}+m_q^2}$ and $\sqrt{2{|Q_uB|}+m_q^2}+\sqrt{2{|Q_dB|}+m_q^2}$. Since the polarization function $\Pi_{\pi_+}$ in the pole equation (\ref{mpmass}) is infrared divergent at these three threshold energies, there are three mass jumps for $\pi_+$ shown in Fig.\ref{fig2}. For magnetic field $eB/m_{\pi}^2=10$ and $20$, the three jump temperatures are respectively $T^+=157,171,197$ MeV and $160,172,187$ MeV.

The mass jumps for $\pi_0$ and $\pi_\pm$ are caused by the discrete energy levels of constituent quarks, although the locations of jumps are different. When the magnetic field is turned off, there will be no more discrete energy levels, the integration in the polarization function $\int d^3{\bf p}/(2E_f-\omega)\sim \int pdp$ becomes finite at the threshold, and the mass jumps disappear automatically, see the upper panel of Fig.\ref{fig2}. It is also necessary to point out that, in hadron models like chiral perturbation theory, linear sigma model and bosonized NJL model where hadrons are taken as elementary particles, the pion mass changes with temperature continuously~\cite{c1,c3,hadron1,hadron2,wang}.

In above calculations, we treated quarks in mean field approximation and mesons in RPA. Since quarks are calculated at mean field level, there are magnetic catalysis effect at both low and high temperatures: the quark mass at $T=0$ shown in Fig.\ref{fig1} and the critical temperature $T_c$ shown in Fig.\ref{fig2} increase with magnetic field. The former agrees with almost all the models~\cite{rev1,rev2,rev3,rev4} but the latter is in contrast with the inverse magnetic catalysis obtained in lattice QCD simulations~\cite{lattice1,lattice2,lattice3,lattice4}. What is the behavior of the pion masses if we take the inverse magnetic catalysis into account? To answer this question in the NJL model, we still take the gap equation (\ref{gap}) but use a magnetic field dependent coupling $G(B)$ by fitting the critical temperature $T_c$ from the lattice simulation~\cite{lattice1}. We numerically checked that, while the location of the mass jump is modified by such a $B$-dependent coupling, the jump itself still survives, since it is a direct consequence of the discrete quark energy level in magnetic field.

\section{Summary}
Pions, including neutral and charged pions, in magnetic field are systematically investigated in the frame of a Pauli-Villars regularized NJL model. Seriously taking into account the breaking of translational invariance for charged particles, we construct mesons in coordinate space with RPA method. By defining an appropriate Fourier-like transformation from coordinate space to the conserved momentum space in Ritus scheme, which is carried out by the eigenfunctions of the magnetized Klein-Gordon equation for mesons, an algebraic equation for meson propagator is obtained in the momentum space. This is a general method and can be extended to other charged mesons like $K$ and $\rho$.

The magnetic field breaks down the isospin symmetry and leads to a mass splitting between neutral and charged pions. At zero temperature, while the pseudo-Goldstone mode $\pi_0$ is only slightly affected by magnetic field, charged pions $\pi_\pm$ become much heavier in the field. This magnetic field dependence is qualitatively consistent with the lattice QCD simulation. At finite temperature, a significant magnetic field effect is the meson mass jump at the Mott transition point due to the discrete quark energy level. For neutral pion, the constituent quark and antiquark can be both at the lowest Landau level, and there is only one jump at the Mott transition point. For charged mesons, however, the two constituents at the lowest Landau level are forbidden, and there are three jumps at the three meson threshold energies for the Mott transition.\\

\noindent {\bf Acknowledgement:}
The work is supported by the NSFC Grant 11775165.

\end{document}